\documentstyle[prl,aps,multicol,epsfig]{revtex}
\begin{document} 
\draft 
\title{Temperature dependent band structure of the Kondo insulator}

\author{C. Gr\"ober and R. Eder}
\address{Institut f\"ur Theoretische Physik, Universit\"at W\"urzburg,
Am Hubland,  97074 W\"urzburg, Germany}
\date{\today}
\maketitle
\begin{abstract}
We present a Qantum Monte Carlo (QMC) study of the temperature dependent
dynamics of the Kondo insulator. Working at the so-called symmetrical point
allows to perform minus-sign free QMC simulations and thus reach 
temperatures of less than 1\% of the conduction electron bandwidth. Study of 
the temperature dependence of the single particle Green's function and
dynamical spin correlation function shows
a surprisingly intricate low temperature band structure and gives evidence
for two characteristic temperatures, which we identify with the
Kondo and coherence temperature, respectively. In particular, the data
show a temperature induced metal-insulator transition at the
coherence temperature.
\end{abstract}
\begin{figure}
\epsfxsize=6.0cm
\vspace{-0.0cm}
\hspace{-0.0cm}\epsfig{file=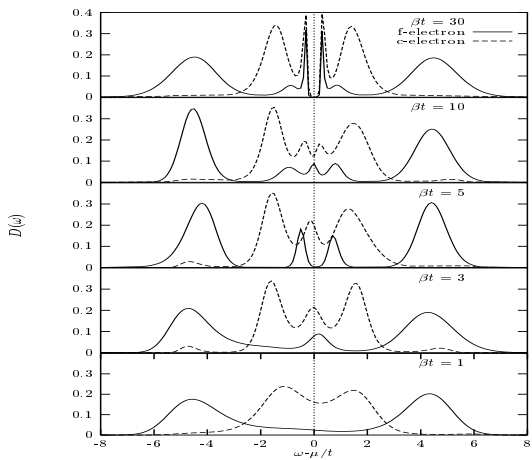,
height=6.0cm,width=7.0cm,angle=0.0}
\vspace{0.5cm}
\narrowtext
\caption[]{Angle integrated spectral density $D(\omega)$ for
an $L$$=$$16$ Kondo lattice for different temperatures.
Parameter values are $U/t$$=$$8$, $V/t$$=$$1$.}
\label{fig1} 
\end{figure}

\begin{figure}
\epsfxsize=6cm
\vspace{-0.0cm}
\hspace{-0.0cm}\epsfig{file=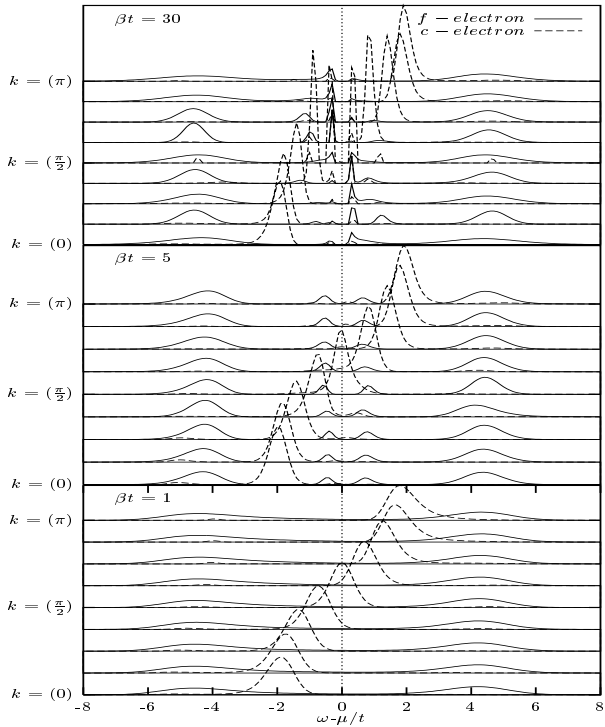,
height=9.5cm,width=8.0cm,angle=0.0}
\vspace{0.5cm}
\narrowtext
\caption[]{Momentum resolved single-particle spectral function
$A(k,\omega)$ at different temperatures. Parameters as in 
Figure \ref{fig1}, the momentum $k$ increases in steps
of $2\pi/L$$=$$\pi/8$ from the bottom of each panel.}
\label{fig2} 
\end{figure}

\begin{figure}
\epsfxsize=6cm
\vspace{-0.0cm}
\hspace{-0.0cm}\epsfig{file=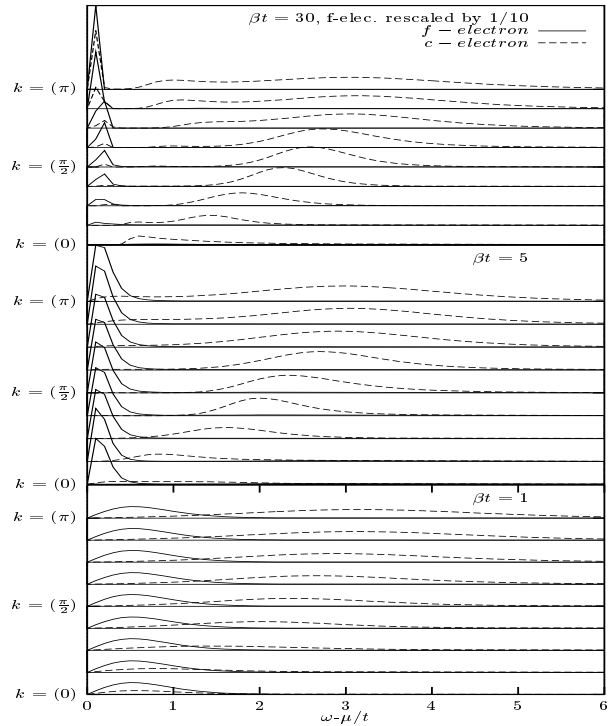,
height=9.5cm,width=8.0cm,angle=0.0}
\vspace{0.5cm}
\narrowtext
\caption[]{Dynamical spin correlation function $S(k,\omega)$
for different temperatures. Parameters are as in Figure \ref{fig1}.}
\label{fig3} 
\end{figure}

 
\begin{multicols}{2}
The theoretical description of the Kondo lattice remains an outstanding 
problem of solid state physics. This model, or variations of it, may be 
viewed as the appropriate one for understanding such intensively investigated 
classes of materials as the heavy electron metals\cite{Stewart,Fulde} and the
Kondo insulators\cite{Fisk}. Experimental results indicate
that the electronic structures of Kondo lattice compounds undergo quite
dramatic changes with 
temperature\cite{Schlesinger}. It is the purpose of the present manuscript
to report a QMC study of the electronic structure of
the so-called Kondo insulator, which shows that this model
indeed undergoes a quite profound change of its unexpectedly
intricate band structure as temperature increases. We are using
a one dimensional (1D) `tight-binding version' of the model
with $L$ unit cells and $2$ orbitals/unit cell:
\begin{eqnarray}
H &=& 
-t\sum_{i,\sigma} 
(c_{i+1,\sigma}^\dagger 
c_{i,\sigma}^{}  + H.c.)
- V \sum_{i,\sigma} (c_{i,\sigma}^\dagger f_{i,\sigma}^{} + H.c.)
\nonumber \\
&-& \epsilon_f \sum_{i,\sigma} n_{i,\sigma}
+ U\sum_{i} f_{i,\uparrow}^\dagger f_{i,\uparrow}^{}
 f_{i,\downarrow}^\dagger f_{i,\downarrow}^{}.
\label{kondo1}
\end{eqnarray}
Here $c_{i,\sigma}^\dagger$  ($f_{i,\sigma}^\dagger$)
creates a conduction electron ($f$-electron)
in cell $i$, $n_{i,\sigma}$$=$$f_{i,\sigma}^\dagger f_{i,\sigma}^{}$.
Throughout we consider the case of `half-filling'
i.e. two electrons/unit cell and, as an important technical point,
we restrict ourselves to the symmetric case,
$\epsilon_f$$=$$U/2$. The latter choice, while probably not
leading to any qualitative change as compared to
other ratios of $\epsilon_f/U$, has the crucial advantage that
at half-filling the model acquires particle-hole symmetry,
i.e. the Hamiltonian becomes invariant under the transformation
$\alpha_{i,\sigma}\rightarrow exp(i\bbox{Q}\cdot \bbox{R}_i)
\alpha_{i,\sigma}^\dagger$, where $\alpha$$=$$c,f$ and
$\bbox{Q}$$=$$(\pi,\pi,...)$ (this holds for bipartite lattices
with only nearest neighbor hopping).
Particle-hole symmetry in turn implies that the QMC-procedure does not
suffer from the notorious `minus-sign problem' anymore, so that
reliable simulations for 
temperatures as low as $\beta t$$=$$30$, corresponding to
$\approx 0.8$\% of the conduction electron
bandwidth, can be performed without problems. This allows to scan
the dynamical correlation functions of the
system over a wide temperature range. Previouslt a QMC study for
the 2 dimensional model at
half-filling was performed by Vekic {\em at al.}\cite{Vekic},
more recently a study of the temperature
dependence of {\em static} susceptibilities for the strong coupling
version of the model has been reported
by Shibata {\em et al.}\cite{Shibata}.\\
It is widely believed that the Kondo lattice has two distinct
characteristic temperatures. At the
Kondo temperature, $T_K$, the $f$-electrons start to form loosely 
bound singlets with the conduction electrons. This manifests itself in
a deviation of the spin susceptibility
from the high-temperature Curie form due to the
`quenching' of the $f$-electron magnetic moment and
an increase of the dc-resistivity due to resonant scattering from
the newly formed low energy bound states\cite{Stewart}. The second 
(and lower) characteristic temperature is 
the coherence temperature, $T_{coh}$, where the local singlets
establish long range coherence
amongst themselves so as to participate in the quasiparticle
bands of a Fermi-liquid like electronic state. 
Experimentally
the coherence temperature is signaled by the onset of a decrease of the dc
resistivity with temperature\cite{Stewart}, and
the formation of 
the `heavy bands' which (judging by the volume of the Fermi surface)
incorporate the $f$-electrons\cite{dhva}. While
there is as yet no experimental proof, one might expect on the basis
of these considerations, that at temperatures above $T_{coh}$ the $f$-electrons
do not participate in the Fermi surface volume, whereas they
do so below. \\
Turning to the Kondo insulator we note that the
electron count for these systems is such that the `Fermi surface'
comprising both, conduction and $f$-electrons, would precisely
fill the Brillouin zone so that the system
is a `nominal' band insulator. If increasing temperature
causes the $f$-electrons to `drop out' of the Fermi surface
volume, this should manifest itself as an insulator-to-metal transition
because the volume of the collapsed Fermi surface is no longer sufficient
to cover the entire Brillouin zone.
Tranferring the above scenario for heavy Fermion metals to the Kondo insulator
one would therefore expect that the system remains
a metal above $T_{coh}$, with a Fermi surface that
excludes the $f$-electrons, and becomes insulating below $T_{coh}$,
when the $f$-electrons participate in the Fermi surface volume
to turn the system into a `nominal' band insulator.
In fact insulator-to-metal transitions
which are induced by temperature\cite{Schlesinger},
or hydrostatic pressure\cite{Cooley} have been observed experimentally.
As will be seen below, our data for the Kondo insulator are
remarkably consistent with such an interpretation.\\
To begin with, we consider the $T$-dependence of the 
single particle spectral function.
This is defined as ($\alpha$$=$$c,f$)
\begin{equation}
A_\alpha(k,\omega) = \frac{1}{Z}
\sum_{\nu,\mu} e^{\beta \omega_{\nu \mu}}
| \langle \nu | \alpha_{k,\sigma} | \mu \rangle |^2 
\delta(\omega - \omega_{\nu \mu}),
\end{equation}
where the sum is over all eigenstates $|\nu\rangle$ 
of $H -\mu N$ in the
grand canonical ensemble, $Z$ denotes the partition function
and $\omega_{\nu \mu}$ the difference of the
energies of the states $\nu$ and $\mu$. Figure
\ref{fig1} shows the angle-integrated 
spectral 
\noindent 
density, $D_\alpha(\omega)$$=$$(1/L)\sum_k A_\alpha(k,\omega)$.
At the lowest temperature, $\beta t$$=$$30$,  $D_{\alpha}(\omega)$ 
shows the behaviour expected for a Kondo insulator:
the $c$-electron density is roughly consistent
with the standard $1D$ tight-binding
density of states, with indications of the two
van-Hove singularities at $\pm 2t$. 
Around $\mu$, however, the spectral density shows a small but unambiguous gap,
which demonstrates the insulating nature of the ground state.
The $f$-like spectral density shows very sharp
low energy peaks at the edges of this gap, as well
as high-intensity `Hubbard bands' at approximately $\pm U/2$.
There are also two weak `side bands' at approximately
$\pm 0.8t$. With the exception of these,
all features are consistent with
exact diagonalization\cite{Tsutsui} and $d\rightarrow \infty$ 
results\cite{Tahvildar}
at $T$$=$$0$. This suggests that the sidebands are already an effect of the
finite temperature and the further development with $T$
confirms this. Inspection of the series $\beta t$$=$$30,10$ 
shows that with increasing $T$ a transfer of spectral weight 
from the low energy peaks at the gap-edges into the sidebands
is taking place. This is accompanied by a narrowing
of the gap, and at $\beta t$$=$$10$ the
gap practically closes; in the spectrum for this temperature the two 
low-energy peaks seem to have collapsed into a single one right at $\mu$ -
we believe that these are in fact still two peaks, which however
are too closely spaced to be resolved. 
Increasing $T$ even further ($\beta t$$=$$5$), the
low energy peaks disappear completely. The $c$-electron
spectral density no longer shows any indication of a gap,
whereas the $f$-like sidebands stay at a relatively
high energy away from $\mu$. 
The extreme low energy states thus have pure $c$-character
(within the resolution of the QMC procedure)
and the system is a metal with a $c$-like Fermi surface.
We therefore interpret the temperature where the gap closes
as the analogue of the coherence temperature.
With increasing $T$ the energy of the sidebands
is lowered and at $\beta t$$=$$3$ they collapse into a single
peak right at $\mu$. This probably indicates
a second transition and at the relatively large value of $\beta t$$=$$1$ 
another reconstruction of the band structure has taken place, namely
the disappearance of the $f$-like sidebands. The upper and lower Hubbard 
band for the
$f$-electrons are now quite broad, and actually the possibility that there
are very low-intensity $f$-like features near
$\mu$ cannot be completely ruled out. However, the overall
trend is quite obviously a strong decrease of the
sidebands. Interpreting the latter as the lattice-analogue
of the Kondo-resonance in the
impurity case\cite{Gunnarson}, the temperature of the second
transition should correspond to the Kondo temperature $T_K$.
In the present case $T_K$ is very high because of
the relatively large value of the $c$-$f$ hybridization,
$V$$=$$t$. We also note that the closing of a gap in the $f$-like
density of states in 2 dimensions was previously found
by Vekic {\em et al.}\cite{Vekic}.\\
To get a more detailed picture, we consider the
$k$-resolved single
particle spectral function, shown in Figure \ref{fig2} for some 
temperatures in between the two transitions.
For $\beta t$$=$$30$ the $c$-electrons show
a standard $\cos(k)$ band, albeit
with a clear gap at $k_F^0$$=$$\pi/2$, the Fermi momentum
of nonhybridized conduction electrons. At this momentum
the band changes its spectral character and `bends over' into
a practically dispersionless $f$-like band,
which can be followed up to $k$$=$$\pi$. This kind of band structure
is familiar from various studies\cite{Tsutsui,Tahvildar,eos}.
The weak and practically dispersionless $f$-like sidebands are at
somewhat higher energy and, at very high energies, the $f$-like Hubbard bands. 
The width of the Hubbard bands seems to depend strongly on
momentum - this is a deficiency of the
QMC and maximum entropy method, which is most accurate
near $\mu$.
As seen in the $k$-integrated spectra, raising the temperature
leads to a transfer of spectral weight
from the `flat band' forming the single-particle gap into
the Kondo resonance-like
sidebands. At $\beta t$$=$$5$, where the gap has closed,
the $c$-electron spectrum shows a very conventional $\cos(k)$-band
with no more indication of any gap. At $k_F^0$ there
is now one symmetric and unsplit peak
right at $\mu$ - as it is required by particle-hole
symmetry for a metallic system. The system thus 
has a 
\noindent 
Fermi surface as expected for unhybridized $c$-electrons, i.e. 
the $f$-electrons indeed have `dropped out' of the
Fermi surface volume. The dispersionless $f$-like sidebands
are at a relatively high energy. The tiny low energy `foot'
seen in some of the $c$-like spectra may indicate
a very weak mixing of the $c$-electrons into the Kondo-resonance
but apparently this no longer leads to a gap.
Rather, the Kondo resonance is now essentially decoupled
from the Fermi surface physics.
Finally, for $\beta t$$=$$1$, even these Kondo resonance-like sidebands
have disappeared and the only $f$-like peaks in the
spectral function are the upper and lower Hubbard bands.
At this high temperature the $f$-electrons do not participate in the
low energy physics at all. The expectation value
of $- V \sum_{i,\sigma} (c_{i,\sigma}^\dagger f_{i,\sigma}^{} + H.c.)$
decreases by $\approx 30$\% between $\beta t$$=$$30$ and
$\beta t$$=$$1$ - while there is
appreciable mixing even at high temperature, this
does obviously not lead to coherent band formation any more.\\
We proceed to the dynamical spin correlation function (SCF),
defined as
\begin{equation}
S(k,\omega) = \frac{1}{Z}
\sum_{\nu,\mu} e^{-\beta E_\mu}
| \langle \nu | S^z_\alpha(k) | \mu \rangle |^2 
\delta(\omega - \omega_{\nu \mu}),
\end{equation}
where $S^z_\alpha(k)$ is (the Fourier transform of)
the $z$-component of the spin-operator for $\alpha$-electrons.
This is shown in Figure \ref{fig3}.
At $\beta t$$=$$30$, the $f$-like SCF shows an intense
branch of low energy excitations with a tiny but clearly resolved
spin wave-like dispersion. The spectral weight of this branch is sharply
peaked at $k$$=$$\pi$, indicating relatively long ranged and
strong antiferromagnetic spin correlations. Fitting the 
equal (imaginary) time $f$-like spin correlation function in real space
to the expression $S(r)$$=$$A( e^{-r/\zeta}$$+$$
e^{-(L-r)/\zeta})$ we obtain the values
$\zeta$$=$$4.61,4.67$ and $2.18$ for $\beta t$$=$$30,20$
and $10$. The dominant feature in the
$c$-like SCF on the other hand, is a free electron-like particle-hole
continuum. Interestingly enough, there is also a
replica of the $f$-electron spin wave branch in the
$c$-like SCF. This shows that at low excitation energies
$c$ and $f$ electrons behave as a single `all-electron fluid',
whose excitations have composite $f$-$c$ character.
The free electron continuum itself does have a gap
of $\approx 0.6t$ at $k$$=$$\pi$ - this corresponds
to approximately 
\noindent 
twice the single particle-gap in 
$A(k,\omega)$\cite{Tsunetsugu}.
We proceed to $\beta t$$=$$5$, where the single-particle gap has closed.
The $f$-like SCF still shows a low-energy peak with finite
excitation energy, which however is practically dispersionless,
both with respect to its energy and with respect to its spectral
weight. In other words, the magnetic
$f$-excitation becomes practically immobile.
The metal-insulator transition thus is related to a drastic
change of the spin correlation function\cite{Vekic} - the question
whether the longer ranged spin correlations below the transition
are the driving force behind the gap formation\cite{Tsunetsugu} or whether
the change of the spin correlations is merely a `byproduct' of the
collapse of the single-particle gap, remains to be clarified.
In any way, the almost completely localized spin dynamics of the
$f$-electrons naturally should lead to a Curie-law for the
static spin susceptibility at temperatures above
$T_{coh}$. A crossover from an activation-gap
dominated susceptibility at low temperatures
and a Curie law at high temperatures has indeed been
observed by Shibata {\em et al.}\cite{Shibata}.
In the $c$-like SCF the
particle-hole continuum persists and the gap near $k$$=$$\pi$
is now very small or zero (the absence of the gap in
the single particle spectrum suggests it to be zero).
There is no more distinguishable peak in the $c$-like SCF which would
correspond to the dispersionless $f$-like spin resonance -
this suggests that the $f$-electrons now are largely
decoupled from the
$c$-like band, as indicated by their non-participation
in the Fermi surface. At the very high
$\beta t$$=$$1$, there is still some (very weak) indication of the
low energy $f$-electron spin resonance, but the intensity is
low and the resonance is now relatively broad. It should be noted that
the relative change of the $f$-like magnetic moment is less than
$2$\% over the entire temperature range we studied - temperature
thus affects only the coupling of these moments to the conduction
electrons.\\ 
For the relatively large value of $V$$=$$1$ the higher of the two
crossover temperatures (i.e. the Kondo temperature) is already 
rather high, $\beta t$$\ge$$1$. Based on the impurity results\cite{Wilson}
one might expect that for smaller $V$ the Kondo temperature
is lower, and to check this, we have performed 
\begin{figure}
\epsfxsize=6cm
\vspace{-0.0cm}
\hspace{-0.0cm}\epsfig{file=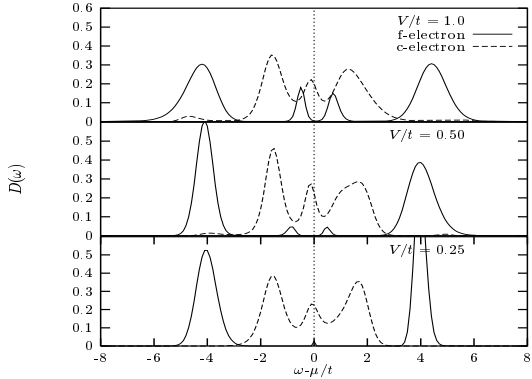,
height=5.0cm,width=7.0cm,angle=0.0}
\vspace{0.5cm}
\narrowtext
\caption[]{Single particle spectral density 
$D(\omega)$ at $\beta t$$=$$5$
for different $V/t$. All other parameters are as in
Figure \ref{fig1}.}
\label{fig4} 
\end{figure}
\noindent 
simulations at fixed $\beta t$$=$$5$, but with variable hybridization.
Figure \ref{fig4} shows the results for the
single particle spectral density. 
The weight of the Kondo resonance-like sidebands for fixed temperature
decreases with $V$, and 
for $V/t$$=$$0.25$ they have disappeared
completely. In the $k$-resolved $c$-electron spectrum (not shown)
there is now only a very sharp
nearest neighbor-hopping band with a clear Fermi level crossing at
$k_F^0$ and the $f$-like spin correlation function shows no more
indication of the low energy resonance - the overall picture is
completely the same as for $V/t$$=$$1$ and $\beta t$$=$$1$, with the
sole exception that
all features are much sharper due to the lower temperature.
Here we do not pursue the issue of the detailed parameter dependence
of the characteristsic temperatures - it is quite
obvious, however, that lower values of $V$
shift the characteristic temperatures of the
system towards lower values, but otherwise
leave the physics unchanged.\\
In summary, we have studied the temperature evolution of
various dynamical correlation functions of the Kondo insulator
and found indications for two distinct electronic crossovers.
At the low temperature crossover the single-particle gap
closes so that the system becomes a metal, the
magnetic correlations on the $f$-sites become localized,
and the `spin gap' closes due to $c$-like spin excitations.
While $c$ and $f$
electrons seem to form a coherent `all-electron fluid'
below the crossover temperature, the $c$ and $f$-like
features in the correlation functions
above this temperature are
decoupled. We therefore interpret this temperature
as the analogue of the coherence temperature in heavy-Fermion metals.
At the high-temperature crossover both, the dispersionless $f$-like
Kondo-resonance in the single-particle spectrum and the
$f$-like low-energy spin excitation disappear. 
The only remaining $f$-like feature in the single particle spectrum
are the high-energy `Hubbard bands', corresponding to the
`undressed' transitions $f^1 \rightarrow f^0$ and $f^1 \rightarrow f^2$.
We therefore interpret this second temperature as the Kondo temperature 
of the system.
 
\end{multicols}
\end{document}